\documentstyle[12pt]{article}

\begin{document}
\title{Anharmonic oscillators energies via artificial
perturbation method}
\author{ Omar Mustafa and Maen Odeh\\
 Department of Physics, Eastern Mediterranean University\\
 G. Magusa, North Cyprus, Mersin 10 - Turkey\\
 email: omustafa.as@mozart.emu.edu.tr\\
 PACS number(s): 03.65.Fd, 03.65.Ge, 03.65.Sq\\
\date{}\\}
\maketitle
\newpage
\begin{abstract}
{\small A new pseudoperturbative ( artificial in nature) methodical
proposal [15] is used to solve for Schr\"odinger equation with a class
of phenomenologically useful and methodically challenging anharmonic
oscillator potentials $V(q)=\alpha_o q^2+\alpha q^4$. The effect of the
[4,5] Pad\'{e} approximant on the leading eigenenergy term is studied.
Comparison with results from numerical ( exact) and several eligible
( approximation) methods is made.}
\end{abstract}
\newpage

\section{Introduction}

Quartic anharmonic interactions continue to remain a focus of attention.
Their Hamiltonian\\
\begin{equation}
H=\frac{p^2}{2m}+\alpha_0 r^2+ \alpha r^4
\end{equation}\\
forms one of the most popular theoretical laboratories for examining the
validity of various approximation techniques and represents a nontrivial
physics. Interest in this model Hamiltonian arises in quantum field theory
and molecular physics [1-6].

Although enormous progress has been made over the years in our understanding
of this Hamiltonian, questions of delicate nature inevitably arise in the
process. The hardest amongst often relate to the existence of the assumed
small expansion parameter and the universality of an adequately attendant
powerful approximation. The implementation of Rayleigh-Schr\"odinger
perturbation theory, or even naive perturbation series, expresses the
eigenvalues as a formal power series in $\alpha$ which is quite often
divergent, or at best asymptotic, for every $\alpha\neq0$. One has therefore
to sum up such series [7-10]. Hence, apparently artificial perturbation recipes have been devised
and shown to be ways to make progress [2,3,11-16].
Without being exhaustive, several eligible methods have been used to
calculate the eigenvalues and eigenfunctions for Hamiltonian (1). Long
lists of these could be found in Ref.s[2,3,8-10,13,17-19].

In this paper we introduce, in section 2, a new analytical
( or, preferably, semianalytical)
perturbation method for solving Schr\"odinger equation. The construction
of which starts with the time-independent one-dimensional form of
Schr\"odinger equation, in $\hbar=m=1$ units,\\
\begin{equation}
\left[-\frac{1}{2}\frac{d^{2}}{dq^{2}}+\frac{l(l+1)}{2q^{2}}+V(q)\right]
\Psi_{n_r,l}(q)=E_{n_r,l}\Psi_{n_r,l}(q),
\end{equation}\\
where $l$ is some quantum number and $n_r$ counts the nodal zeros in
$\Psi_{n_r,l}(q)$. The symmetry of an attendant problem obviously
manifests  the admissibility of the quantum number $l$: In one-dimension (1D),
$l$ specifies parity, $(-1)^{l+1}$, with the permissible values -1
and/or 0 ( even and/or odd parity, respectively) where
$q=x\in(-\infty,\infty)$. For two-dimensional (2D) cylindrically symmetric
Schr\"odinger equation  one sets $l=|m|-1/2$, where
m is the magnetic quantum number and $q=(x^2+y^2)^{1/2}\in(0,\infty)$.
Finally, for three-dimensional (3D) spherically symmetric Schr\"odinger
equation, $l$ denotes the angular momentum quantum number with
$q=(x^2+y^2+z^2)^{1/2}\in(0,\infty)$.

We shall focus our attention, in section 3, on
1D and 3D problems and consider, for the sake of diversity; (i) 3D
anharmonic oscillators $V(r)=r^2/2+r^4/2$ with $n_r=0$ and
$l=0,1,2,5,10,50$, (ii) 3D ground state, or equivalently 1D first excited
( odd-parity) state, for anharmonic oscillators $V(q)=q^2/2+\alpha q^4$
over a wide range of anharmonicities ( i.e.; $\alpha=0.002$ to
$\alpha=20000$), and (iii) 3D single-well anharmonic oscillator ground state,
or equivalently 1D double-well anharmonic oscillator first excited state, for
$V(q)=-aq^2/2+q^4/2$ at various well depths ( i.e.; $a=1,5,10,15,25,50,100$).
For the sake of comparison, we use results from exact numerical methods
reported in [2,5], the best estimation of the phase-integral method (PIM)
[5], an open perturbation technique [2], and a perturbative-variational
method (PVM) [6]. Section 4 is reserved for concluding remarks.

\section{The Method}

Our methodical proposal uses $1/\bar{l}$ as a perturbation expansion
parameter, where $\bar{l}=l-\beta$ and $\beta$ is a suitable shift mainly
introduced to avoid the trivial case $l=0$. Hence, hereafter, it will be
referred to as the pseudoperturbative ( artificial in nature) shifted-$l$
expansion technique (PSLET). Equation (2) thus becomes\\

\begin{equation}
\left\{-\frac{1}{2}\frac{d^{2}}{dq^{2}}+\tilde{V}(q)\right\}
\Psi_{n_r,l} (q)=E_{n_r,l}\Psi_{n_r,l}(q),
\end{equation}\\
\begin{equation}
\tilde{V}(q)=\frac{\bar{l}^{2}+(2\beta+1)\bar{l}
+\beta(\beta+1)}{2q^{2}}+\frac{\bar{l}^2}{Q}V(q).
\end{equation}\\
Herein, it should be noted that Q is a constant that scales the potential 
$V(q)$ at large - $l$ limit and is set, for any specific choice of $l$
and $n_r$, equal to $\bar{l}^2$ at the end of the calculations [11,16].
And, $\beta$ is to be determined in the sequel.

PSLET procedure begins with shifting the origin of the
coordinate through\\
\begin{equation}
x=\bar{l}^{1/2}(q-q_{o})/q_{o},
\end{equation}\\
where $q_{o}$ is currently an arbitrary point to perform Taylor expansions
about, with its particular value to be determined. Expansions about
this point, $x=0$ (i.e. $q=q_o$), yield\\
\begin{equation}
\frac{1}{q^{2}}=\sum^{\infty}_{n=0} (-1)^{n} \frac{(n+1)}{q_{o}^{2}}
 x^{n}\bar{l}^{-n/2},
\end{equation}\\
\begin{equation}
V(x(q))=\sum^{\infty}_{n=0}\left(\frac{d^{n}V(q_{o})}{dq_{o}^{n}}\right)
\frac{(q_{o}x)^{n}}{n!}\bar{l}^{-n/2}.
\end{equation}\\
Obviously, the expansions in (6) and (7) center the problem at an
arbitrary point $q_o$ and the derivatives, in effect, contain
information not only at $q_o$ but also at any point on $q$-axis, in
accordance with Taylor's theorem. Also it should be mentioned here
that the scaled coordinate, equation (5), has no effect on the energy
eigenvalues, which are coordinate - independent. It just facilitates
the calculations of both the energy eigenvalues and eigenfunctions.
It is also convenient to expand $E$ as\\
\begin{equation}
E_{n_r,l}=\sum^{\infty}_{n=-2}E_{n_r,l}^{(n)}\bar{l}^{-n}.
\end{equation}\\
Equation (3) thus becomes\\
\begin{equation}
\left[-\frac{1}{2}\frac{d^{2}}{dx^{2}}+\frac{q_{o}^{2}}{\bar{l}}
\tilde{V}(x(q))\right]
\Psi_{n_r,l}(x)=\frac{q_{o}^2}{\bar{l}}E_{n_r,l}\Psi_{n_r,l}(x),
\end{equation}\\
with\\
\begin{eqnarray}
\frac{q_o^2}{\bar{l}}\tilde{V}(x(q))&=&q_o^2\bar{l}
\left[\frac{1}{2q_o^2}+\frac{V(q_o)}{Q}\right]
+\bar{l}^{1/2}\left[-x+\frac{V^{'}(q_o)q_o^3 x}{Q}\right]\nonumber\\
&+&\left[\frac{3}{2}x^2+\frac{V^{''}(q_o) q_o^4 x^2}{2Q}\right]
+(2\beta+1)\sum^{\infty}_{n=1}(-1)^n \frac{(n+1)}{2}x^n \bar{l}^{-n/2}
\nonumber\\
&+&q_o^2\sum^{\infty}_{n=3}\left[(-1)^n \frac{(n+1)}{2q_o^2}x^n
+\left(\frac{d^n V(q_o)}{dq_o^n}\right)\frac{(q_o x)^n}{n! Q}\right]
\bar{l}^{-(n-2)/2}\nonumber\\
&+&\beta(\beta+1)\sum^{\infty}_{n=0}(-1)^n\frac{(n+1)}{2}x^n
\bar{l}^{-(n+2)/2}+\frac{(2\beta+1)}{2},
\end{eqnarray}\\
where the prime of $V(q_o)$ denotes derivative with respect to $q_o$.
Equation (9) is exactly of the type of Schr\"odinger equation 
for one - dimensional anharmonic oscillator\\
\begin{equation}
\left[-\frac{1}{2}\frac{d^2}{dx^2}+\frac{1}{2}w^2 x^2 +\varepsilon_o
+P(x)\right]X_{n_r}(x)=\lambda_{n_r}X_{n_r}(x),
\end{equation}\\
where $P(x)$ is a perturbation - like term and $\varepsilon_o$ is a 
constant. A simple comparison between Eqs.(9), (10) and (11) implies\\
\begin{equation}
\varepsilon_o =\bar{l}\left[\frac{1}{2}+\frac{q_o^2 V(q_o)}{Q}\right]
+\frac{2\beta+1}{2}+\frac{\beta(\beta+1)}{2\bar{l}},
\end{equation}\\
\begin{eqnarray}
\lambda_{n_{r}}&=&\bar{l}\left[\frac{1}{2}+\frac{q_o^2 V(q_o)}{Q}\right]
+\left[\frac{2\beta+1}{2}+(n_r+\frac{1}{2})w\right]\nonumber\\
&+&\frac{1}{\bar{l}}\left[\frac{\beta(\beta+1)}{2}+\lambda_{n_{r}}^{(0)}\right]
+\sum^{\infty}_{n=2}\lambda_{n_{r}}^{(n-1)}\bar{l}^{-n},
\end{eqnarray}\\
and\\
\begin{equation}
\lambda_{n_{r}} = q_o^2 \sum^{\infty}_{n=-2} E_{n_r,l}^{(n)}
\bar{l}^{-(n+1)},
\end{equation}\\
Equations (13) and (14) yield\\
\begin{equation}
E_{n_r,l}^{(-2)}=\frac{1}{2q_o^2}+\frac{V(q_o)}{Q}
\end{equation}\\
\begin{equation}
E_{n_r,l}^{(-1)}=\frac{1}{q_o^2}\left[\frac{2\beta+1}{2}
+(n_r +\frac{1}{2})w\right]
\end{equation}\\
\begin{equation}
E_{n_r,l}^{(0)}=\frac{1}{q_o^2}\left[ \frac{\beta(\beta+1)}{2}
+\lambda_{n_r}^{(0)}\right]
\end{equation}\\
\begin{equation}
E_{n_r,l}^{(n)}=\lambda_{n_r}^{(n)}/q_o^2  ~~;~~~~n \geq 1.
\end{equation}\\
Here $q_o$ is chosen to minimize $E_{n_r,l}^{(-2)}$, i. e.\\
\begin{equation}
\frac{dE_{n_r,l}^{(-2)}}{dq_o}=0~~~~
and~~~~\frac{d^2 E_{n_r,l}^{(-2)}}{dq_o^2}>0.
\end{equation}\\
Hereby, $V(q)$ is assumed to be well behaved so that $E^{(-2)}$ has
a minimum $q_o$ and there are well - defined bound - states.
Equation (19) in turn gives, with $\bar{l}=\sqrt{Q}$,\\
\begin{equation}
l-\beta=\sqrt{q_{o}^{3}V^{'}(q_{o})}.
\end{equation}\\
Consequently, the second term in Eq.(10) vanishes and the first term adds 
a constant to the energy eigenvalues. It should be noted that energy term
$\bar{l}^2E_{n_r,l}^{(-2)}$ has its counterpart in classical
mechanics. It corresponds roughly to the energy of a classical particle
with angular momentum $L_z$=$\bar{l}$  executing circular motion  of 
radius $q_o$ in the potential $V(q_o)$. This term thus identifies the 
leading - order approximation, to all eigenvalues, as a classical 
approximation and the higher - order corrections as quantum fluctuations
around the minimum $q_o$, organized in inverse powers of $\bar{l}$.
The next leading correction to the energy series, $\bar{l}E_{n_r,l}^{(-1)}$,
consists of a constant term and the exact eigenvalues of the unperturbed
harmonic oscillator potential $w^2x^2/2$. 
The shifting parameter $\beta$ is determined by choosing
$\bar{l}E_{n_r,l}^{(-1)}$=0. This choice is physically motivated. It requires
not only the agreements between PSLET eigenvalues and the exact known ones for
the harmonic oscillator and Coulomb potentials but also between the
eigenfunctions.  Hence\\
\begin{equation}
\beta=-\left[\frac{1}{2}+(n_{r}+\frac{1}{2})w\right],
\end{equation}\\
where\\
\begin{equation}
w=\sqrt{3+\frac{q_o V^{''}(q_o)}{V^{'}(q_o)}}.
\end{equation}\\

Then equation (10) reduces to\\
\begin{equation}
\frac{q_o^2}{\bar{l}}\tilde{V}(x(q))=
q_o^2\bar{l}\left[\frac{1}{2q_o^2}+\frac{V(q_o)}{Q}\right]+
\sum^{\infty}_{n=0} v^{(n)}(x) \bar{l}^{-n/2},
\end{equation}\\
where\\
\begin{equation}
v^{(0)}(x)=\frac{1}{2}w^2 x^2 + \frac{2\beta+1}{2},
\end{equation}\\
\begin{equation}
v^{(1)}(x)=-(2\beta+1) x - 2x^3 + \frac{q_o^5 V^{'''}(q_o)}{6 Q} x^3,
\end{equation}\\
and for $n \geq 2$\\
\begin{eqnarray}
v^{(n)}(x)&=&(-1)^n (2\beta+1) \frac{(n+1)}{2} x^n
+ (-1)^{n} \frac{\beta(\beta+1)}{2} (n-1) x^{(n-2)}\nonumber\\
&+& \left[(-1)^{n} \frac{(n+3)}{2}
+ \frac{q_o^{(n+4)}}{Q(n+2)!} \frac{d^{n+2} V(q_o)}{dq_o^{n+2}}\right]
x^{n+2}.
\end{eqnarray}\\
Equation (9) thus becomes\\
\begin{eqnarray}
&&\left[-\frac{1}{2}\frac{d^2}{dx^2} + \sum^{\infty}_{n=0} v^{(n)}
\bar{l}^{-n/2}\right]\Psi_{n_r,l} (x)= \nonumber\\
&& \left[\frac{1}{\bar{l}}\left(\frac{\beta(\beta+1)}{2}
+\lambda_{n_r}^{(0)}\right) 
+ \sum^{\infty}_{n=2} \lambda_{n_r}^{(n-1)}
\bar{l}^{-n} \right] \Psi_{n_r,l}(x).
\end{eqnarray}\\

Up to this point, one would conclude that the above procedure is nothing
but an imitation of the eminent shifted large-N expansion (SLNT)
[12,14,16,20-22]. However, because of the limited capability of SLNT
in handling large-order corrections via the standard Rayleigh-Schr\"odinger
perturbation theory, only low-order corrections have been reported,
sacrificing in effect its preciseness. Therefore, one should seek
for an alternative and proceed by setting the nodeless, $n_r = 0$,
wave functions as \\
\begin{equation}
\Psi_{0,l}(x(q)) = exp(U_{0,l}(x)).
\end{equation}\\
In turn, equation (27) readily transforms into the
following Riccati equation [2,3, and references therein]:\\
\begin{eqnarray}
-\frac{1}{2}[ U^{''}(x)+U^{'}(x)U^{'}(x)]
+\sum^{\infty}_{n=0} v^{(n)}(x) \bar{l}^{-n/2}
&=&\frac{1}{\bar{l}} \left( \frac{\beta(\beta+1)}{2}
+ \lambda_{0}^{(0)}\right)\nonumber\\
&&+\sum^{\infty}_{n=2} \lambda_{0}^{(n-1)} \bar{l}^{-n}.
\end{eqnarray}\\
Hereafter, we shall use $U(x)$ instead of $U_{0,l}(x)$ for simplicity,
and the prime of $U(x)$ denotes derivative with respect to $x$. It is
evident that this equation admits solution of the form \\
\begin{equation}
U^{'}(x)=\sum^{\infty}_{n=0} U^{(n)}(x) \bar{l}^{-n/2}
+\sum^{\infty}_{n=0} G^{(n)}(x) \bar{l}^{-(n+1)/2},
\end{equation}\\
where\\
\begin{equation}
U^{(n)}(x)=\sum^{n+1}_{m=0} D_{m,n} x^{2m-1} ~~~~;~~~D_{0,n}=0,
\end{equation}\\
\begin{equation}
G^{(n)}(x)=\sum^{n+1}_{m=0} C_{m,n} x^{2m}.
\end{equation}\\
Substituting equations (30) - (32) into equation (29) implies\\
\begin{eqnarray}
&-&\frac{1}{2} \sum^{\infty}_{n=0}\left[U^{(n)^{'}} \bar{l}^{-n/2}
+ G^{(n)^{'}} \bar{l}^{-(n+1)/2}\right] \nonumber\\
&-&\frac{1}{2} \sum^{\infty}_{n=0} \sum^{\infty}_{p=0}
\left[ U^{(n)}U^{(p)} \bar{l}^{-(n+p)/2}
+G^{(n)}G^{(p)} \bar{l}^{-(n+p+2)/2}
+2 U^{(n)}G^{(p)} \bar{l}^{-(n+p+1)/2}\right]\nonumber\\
&+&\sum^{\infty}_{n=0}v^{(n)} \bar{l}^{-n/2}
=\frac{1}{\bar{l}}\left(\frac{\beta(\beta+1)}{2}+\lambda_{0}^{(0)}\right)
+\sum^{\infty}_{n=2} \lambda_{0}^{(n-1)} \bar{l}^{-n},
\end{eqnarray}\\
where primes of $U^{(n)}(x)$ and $G^{(n)}(x)$ denote derivatives
with respect to $x$. Equating the coefficients of the same powers of
$\bar{l}$ and $x$, respectively, ( of course the other way around would 
work equally well) one obtains\\
\begin{equation}
-\frac{1}{2}U^{(0)^{'}} - \frac{1}{2}  U^{(0)} U^{(0)} + v^{(0)} = 0,
\end{equation}\\
\begin{equation}
U^{(0)^{'}}(x) = D_{1,0} ~~~;~~~~D_{1,0}=-w,
\end{equation}\\
and integration over $x$ yields\\
\begin{equation}
U^{(0)}(x)=-wx.
\end{equation}\\
Similarly,\\
\begin{equation}
-\frac{1}{2}[U^{(1)^{'}} + G^{(0)^{'}}] - U^{(0)}U^{(1)} - U^{(0)}G^{(0)}
+v^{(1)}=0,
\end{equation}\\
\begin{equation}
U^{(1)}(x)=0,
\end{equation}\\
\begin{equation}
G^{(0)}(x)=C_{0,0}+C_{1,0}x^2,
\end{equation}\\
\begin{equation}
C_{1,0}=-\frac{B_{1}}{w},
\end{equation}\\
\begin{equation}
C_{0,0}=\frac{1}{w}(C_{1,0}+2\beta+1),
\end{equation}\\
\begin{equation}
B_{1}=-2+\frac{q_o^5}{6Q}\frac{d^3 V(q_o)}{dq_o^3},
\end{equation}\\
\begin{eqnarray}
&&-\frac{1}{2}[U^{(2)^{'}} + G^{(1)^{'}}]
- \frac{1}{2}\sum^{2}_{n=0}U^{(n)}U^{(2-n)}-\frac{1}{2}G^{(0)}G^{(0)}
\nonumber\\
&&-\sum^{1}_{n=0}U^{(n)}G^{(1-n)} + v^{(2)}
= \frac{\beta(\beta+1)}{2} + \lambda_{0}^{(0)},
\end{eqnarray}\\
\begin{equation}
U^{(2)}(x)=D_{1,2}x + D_{2,2}x^3,
\end{equation}\\
\begin{equation}
G^{(1)}(x)=0,
\end{equation}\\
\begin{equation}
D_{2,2}=\frac{1}{w}(\frac{C_{1,0}^2}{2}-B_{2})
\end{equation}\\
\begin{equation}
D_{1,2}=\frac{1}{w}(\frac{3}{2}D_{2,2}+C_{0,0}C_{1,0}
-\frac{3}{2}(2\beta+1)),
\end{equation}\\
\begin{equation}
B_{2}=\frac{5}{2}+\frac{q_o^6}{24Q}\frac{d^4V(q_o)}{dq_o^4},
\end{equation}\\
\begin{equation}
\lambda_{0}^{(0)} = -\frac{1}{2}(D_{1,2}+C_{0,0}^2).
\end{equation}\\
and so on. Thus, one can calculate the energy 
eigenvalue and the eigenfunctions from the knowledge of $C_{m,n}$
and $D_{m,n}$ in a hierarchical manner.
Nevertheless, the procedure just described is suitable for systematic calculations
using software packages (such as MATHEMATICA, MAPLE, or REDUCE) to determine
the energy eigenvalue and eigenfunction corrections up to any order of the
pseudoperturbation series. 

Although the energy series, Eq.(8), could appear
divergent, or, at best, asymptotic for small $\bar{l}$, one can still 
calculate the eigenenergies to a very good accuracy by forming the 
sophisticated [N,M+1] Pade' approximation\\
\begin{center}
$P_{N}^{M+1}(1/\bar{l})=(P_0+P_1/\bar{l}+\cdots+P_M/\bar{l}^M)/
(1+q_1/\bar{l}+\cdots+q_N/\bar{l}^N)$
\end{center}
to the energy series [23]. The energy series, Eq.(8), is calculated up to
$E_{0,l}^{(8)}/\bar{l}^8$ by
\begin{equation}
E_{0,l}=\bar{l}^{2}E_{0,l}^{(-2)}+E_{0,l}^{(0)}+\cdots
+E_{0,l}^{(8)}/\bar{l}^8+O(1/\bar{l}^{9}),
\end{equation}\\
and with the $P_{4}^{5}(1/\bar{l})$ Pade' approximant it becomes\\
\begin{equation}
E_{0,l}[4,5]=\bar{l}^{2}E_{0,l}^{(-2)}+P_{4}^{5}(1/\bar{l}).
\end{equation}\\

\section{Quartic anharmonic interactions}

Let us consider the phenomenologically useful and methodically challenging
quartic anharmonic interactions\\
\begin{equation}
V(q)=\alpha_o q^2 + \alpha q^4
\end{equation}\\
of Hamiltonian (1). Equation (22) then reads\\
\begin{equation}
w=\sqrt{\frac{8\alpha_o q_o+24\alpha q_o^3}{2\alpha_o q_o+4\alpha q_o^3}},
\end{equation}\\
and Eq.(20) yields\\
\begin{equation}
l+\frac{1}{2}\left(1+\sqrt{\frac{8\alpha_o q_o+24\alpha q_o^3}
{2\alpha_o q_o+4\alpha q_o^3}}\right)
=q_o^2 \sqrt{2\alpha_o+4\alpha q_o^2}.
\end{equation}\\
In the absence of a closed form solution for $q_o$ in (54), one should
appeal to some software packages ( MAPLE is used here) to resolve this
issue. Of course there is always more than one root for (54). However, the
symmetry of the problem in hand along with Eq.(19) would single out one
eligible root $q_o$ as a minimum of $E^{(-2)}$. Once $q_o$ is determined
the coefficients $C_{m,n}$ and $D_{m,n}$ are obtained in a sequential manner.
Consequently, the eigenvalues, Eq.(50), and eigenfunctions, Eqs.(30)-(32),
are calculated in the same batch for each value of $\alpha_o$, $\alpha$,
and $l$.

Our results ( tables 1-3) are obtained from the first eleven terms of our
energy series (50). Also, the effect of the [4,5] Pad\'{e} approximant on
the leading term $\bar{l}^2E^{(-2)}$ is reported as E[4,5]. In table 1 we
list our results along with the exact numerical ones and the (best estimated)
eigenvalues obtained from the fifth-order phase-integral method (PIM)
reported by Lakshmanen et al. [5]. Obviously, our results compare excellently
with the exact numerical ones and surpass those from PIM. Whilst the [4,5]
Pad\'{e} approximant had no dramatic effect on the energy eigenvalues for
$l=0$, it had no effect on the energy eigenvalues for $l\geq1$. A common
feature between PSLET and PIM is well pronounced here; the precession of both
methods increases as $l$ increases.

Again we proceed with the theoretical laboratory (52) and examine the validity
of PSLET over a wide range of anharmonicities for $V(q)=q^2/2+\alpha q^4$.
In table 2 we list our results for the three-dimensional (3D) ground states
energies, or equivalently for the one-dimensional (1D) first excited state
energies. The results of Bessis and Bessis [2], via an open perturbation
recipe, and the exact ones [24], using Bargman representation, are also
displayed. Clearly and satisfactorily, the trend of the exact values of
the energies is reproduced.

Finally, we consider the ground state energies of the 3D single-well, or
equivalently the first excited state energies of the 1D double-well,
potentials $V(q)=-aq^2/2+q^4/2$. We compare our results ( table 3) with
those obtained by Saavedra and Buendia [6] via a perturbative-variational
method (PVM). They are in excellent agreement not only with the PVM but
also with the hypervirial perturbation method [25], especially for deep
wells.

\section{Concluding remarks}

The method (PSLET) just described is conceptually sound. It avoids
 troublesome questions
such as those pertaining to the nature of small-parameter expansions, the
trend of convergence to the exact numerical values, the utility in
calculating the eigenvalues and eigenfunctions (in one batch) to
sufficiently heigher-orders, and the applicability to a wide rang of
potentials. Provided that the latter is analytic and give rise to one
minimum of $E^{(-2)}$ and an infinite number of bound states.

On the computational and practical methodology sides, PSLET comes in quite
handy and very accurate numerical results are obtained. Nevertheless,
if greater accuracy is in demand, another suitable criterion for choosing
the value of the shift $\beta$, reported in [13,25], is also feasible.
However, one would always be interested, for practical exploratory purposes,
in the conventional wisdom of perturbation prescriptions that only a few
terms of a "most useful" perturbation series reveal the important features
of the solution before a state of exhaustion is reached. Our method indeed
belongs to this category where the results of the illustrative
challenging examples used bear this out.

On the other hand, asymptotic wavefunctions emerge in our procedure from
the knowledge of $C_{m,n}$ and $D_{m,n}$ to study, for example,
electronic transitions and multiphoton emission occurring in atomic
systems. Such studies already
lie beyond the scope of our present methodical proposal.

\newpage

\begin{table}
\begin{center}
\caption{ Eigenvalues from the fifth-order phase-integral method $E_{PIM}$
[5], the pseudoperturbative shifted-$l$ expansion technique $E_{PSLET}$,
the effect of the [4,5] Pad\'{e} approximant on our leading energy term
E[4,5], and from the exact numerical calculations [5] for the
three-dimensional anharmonic oscillator $V(r)=\frac{1}{2}r^{2}
+\frac{1}{2} r^{4}$, with $n_r=0$.}
\vspace{1cm}
\begin{tabular}{|l|l|l|l|}
\hline\hline
 & $l$=0 & $l$=1 & $l$=2 \\
\hline
$E_{PIM}$     & 2.324 83   & 4.190 26  & 6.242 80\\
$E_{PSLET}$   & 2.324 40   & 4.190 17  & 6.242 78\\
E[4,5]        & 2.324 41   & 4.190 17  & 6.242 78\\
$E_{exact}$   & 2.324 41   & 4.190 17  & 6.242 78\\ \hline
 & $l$=5 & $l$=10 & $l$=50 \\ \hline\hline
$E_{PIM}$   & 13.264 459 9 & 27.092 492 362 & 187.529 708 014 021 \\
$E_{PSLET}$ & 13.264 458 8 & 27.092 492 304 & 187.529 708 014 0025 \\
E[4,5]      & 13.264 458 8 & 27.092 492 304 & 187.529 708 014 0025 \\
$E_{exact}$ & 13.264 458 8 & 27.092 492 305 & 187.529 708 014 003 \\ \hline
\end{tabular}
\end{center}
\end{table}

\newpage

\begin{table}
\begin{center}
\caption{ Three-dimensional ground state energies or equivalently
one-dimensional first excited state energies for 
$V(q)=\frac{q^{2}}{2}+\alpha q^{4}$. $E_{BB}$ denotes Bessis and Bessis
results [2] and the exact ones $E_{exact}$, reported therein, for
different anharmonicities.}
\vspace{1cm}
\begin{tabular}{|l|l|l|l|l|}
\hline\hline
$\alpha $ & $E_{PSLET}$ & $E[4,5]$ & $E_{BB}$ & $E_{exact}$ \\
\hline
0.002 & 1.507 41940 & 1.507 41940  & 1.507 4194 & 1.507 41939 \\ 
0.006 & 1.521 80570 & 1.521 80570 & 1.521 8057 & 1.521 80565 \\ 
0.01  & 1.535 64844 & 1.535 64846 & 1.535 6483 & 1.535 64828 \\ 
0.05  & 1.653 439   & 1.653 439   & 1.653 441  & 1.653 43601 \\ 
0.1   & 1.769 512   & 1.769 625   & 1.769 529  & 1.769 50264 \\ 
0.3   & 2.094 678   & 2.094 640   & 2.094 795  & 2.094 64199 \\ 
0.5   & 2.324 401   & 2.324 407   & 2.324 661  & 2.324 40635 \\ 
0.7   & 2.509 16    & 2.509 23    & 2.509 56   & 2.509 22810 \\ 
1     & 2.737 73    & 2.737 91    & 2.738 32   & 2.737 89227 \\ 
2     & 3.292 48    & 3.292 94    & 3.293 50   & 3.292 86782 \\ 
50    & 8.913 21    & 8.916 61    & 8.917 41   & 8.915 09636 \\ 
200   & 14.056 17   & 14.062 53   & 14.062 96  & 14.059 2268 \\ 
1000  & 23.966 93   & 23.978 93   & 23.978 63  & 23.972 2061 \\ 
8000  & 47.880 19   & 47.890 95   & 47.903 66  & 47.890 7687 \\ 
20000 & 64.972 32   & 65.006 64   & 65.004 18  & 64.986 6757\\
\hline
\end{tabular}
\end{center}
\end{table}

\newpage

\begin{table}
\begin{center}
\caption{ Three-dimensional ground state energies or equivalently
one-dimensional first excited state energies for $V(q)=-a q^{2}/2+q^{4}/2$.
$E_{PVM}$ represents the results from perturbative-variational method [6].}
\vspace{1cm}
\begin{tabular}{|c|c|c|c|}
\hline\hline
$a$ & $E_{PVM}$ & E$_{PSLET}$ & E[4,5] 
\\ \hline
1 & 2.834 5 & 2.835 3 & 2.834 4 \\ 
5 & -3.250 68 & -3.250 85 & -3.250 84 \\ 
10 & -20.633 55 & -25.633 69 & -20.633 50 \\ 
15 & -50.841 387 & -50.841 42 & -50.841 42 \\ 
25 & -149.219 456 & -149.219 454 & -149.219 454 \\ 
50 & -615.020 090 9 & -615.020 091 0 & -615.020 091 0 \\ 
100& -2845.867 880 34 & -2485.867 880 337 & -2485.867 880 337 \\
\hline
\end{tabular}
\end{center}
\end{table}
\end{document}